\documentstyle[12pt]{article}
\begin{document}
\begin{flushright}
INR preprint 0974/98 \\
March  1998  \\
\end{flushright}
\vspace*{1.cm}
\begin{center}
{ \large \bf New Algebra of Local Symmetries for Regge Limit of Yang--Mills 
Theories } \\
\vspace*{1.cm}
{\large  Victor A. Matveev and Grigorii B. Pivovarov }\\
\vspace*{0.3cm}
Institute for Nuclear
Research, 117312 Moscow, Russia \\
\end{center}
\vspace{1cm}
\begin{center}
{\large \bf Abstract}
\end{center}
Local effective action is derived to describe Regge asymptotic of Yang--Mills 
theories. Local symmetries of the effective action originating from the 
gauge symmetry of the underlying Yang--Mills theory are studied. 
Multicomponent effective action is introduced to express the symmetry 
transformations as field transformations. The algebra of these symmetries
is decomposed onto a semi-direct sum of commutative algebras and four copies
of the gauge algebra of the underlying Yang--Mills theory. Possibility of 
existence of solitons corresponding to the commutative subalgebra of the 
symmetry algebra is mentioned.

\vspace*{1.5cm}


\newpage

It is well acknowledged that construction of an effective
field theory which would describe the Regge limit of nonabelian
gauge theories is an important problem. Various approaches to such
a construction along with theoretical and phenomenological
motivations see in \cite{Ver,Ar,Lip,KP}.

Here we develop further the approach of \cite{KP}. Its core is to consider 
excitations of the gauge field  moving fast rightward along a space axis $z$
as excitations of a separate field $A^R$ of the effective theory. Analogously,
an independent field  $A^L$ is introduced to describe excitations of the 
underlying theory which are fast left-movers. It was noted in \cite{KP} 
that there exists a finite effective action describing the dynamics of 
these fields in the limit of infinite energy of the relative motion of the 
right- and left-movers. It was also noted in \cite{KP} that the effective 
action admits a multiplicative infrared renormalization.
Corresponding renormgroup reproduces in the case of gluodynamics the known 
Regge trajectory of the Reggeized gluon. 

To move beyond perturbation theory in the resummation of the energy logarithms
of gauge theories, it is crucial to understand the symmetry behind the
multiplicative renormalizability
established in \cite{KP} in the one-loop approximation. It is also important
to understand this symmetry to extend the set of ``good'' (renormilizable)
field theories which can be used to describe the nature. Below we construct 
an algebra of local symmetries of an effective Regge theory which may be useful
in nonperturbative resumation of the energy logarithms and also provides a 
new example of nontrivial local symmetry.  

The first step is to derive a local effective action. The effective action
of \cite{KP} is nonlocal and is a result of integrating out some fields
of the local effective theory we introduce below. It is obtained from
the underlying Yang--Mills theory in two steps. (In what follows we
assume the gauge group of the underlying theory to be $SU(N)$. Generalization 
to arbitrary gauge group is straightforward. We also reduce our consideration 
to a pure Yang--Mills theory without matter fields.) 

The first step connects the underlying Yang--Mills theory
with an extended theory which is equivalent to the Yang--Mills theory but
has an extended field content. Namely, the extended theory
is a theory of four vector gauge fields---$S$(oft), $H$(ard), $R$(ight), and
$L$(eft). Its action is
\begin{equation}
\label{ext}
S_{ext}(\{A^\Gamma\})=S_{YM}(\sum_\Gamma P_\Gamma A^\Gamma),
\end{equation}
where $\{P_\Gamma\},\, \Gamma=\{S,H,L,R\}$ is a complete set of orthogonal 
projectors acting on the fields. For any such a set, vacuum expectations
of products of the fields $P_\Gamma A^\Gamma$ calculated with the action
(\ref{ext}) coincide with the vacuum expectations of the corresponding
products of $P_\Gamma A$, where $A$ is the gauge field, calculated in the 
framework of the underlying Yang--Mills theory:
$$
\langle \Pi_i P_{\Gamma_i}A^{\Gamma_i}\rangle^{S_{ext}}=
\langle \Pi_i P_{\Gamma_i} A\rangle^{S_{YM}}.
$$
The projectors of the particular set we need act on a Fourier transformed
field $\Phi(k)$ as follows:
\begin{eqnarray}
\label{pro}
P_S\Phi(k)=\theta(|k_+|<\mu)\theta(|k_-|<\mu)\Phi(k),\,
P_H\Phi(k)=\theta(|k_+|>\mu)\theta(|k_-|>\mu)\Phi(k),\nonumber\\
P_R\Phi(k)=\theta(|k_+|>\mu)\theta(|k_-|<\mu)\Phi(k),\,
P_L\Phi(k)=\theta(|k_+|<\mu)\theta(|k_-|>\mu)\Phi(k),
\end{eqnarray}
where $\theta(condition)$ equals to unit if $condition$ is fulfilled
and vanishes otherwise. $\mu$ above is an arbitrary scale which will not
enter any physical prediction by renormgroup invariance, and $k_\pm$
are the light-cone components of the momentum, $k_\pm=k_0\pm k_z$. 
Now the labels
$\Gamma$ make sense: $S$(oft) field describes the gluon modes of both light 
cone components of momentum small, etc. 

The last, second step in the derivation of the effective action is to drop out 
some vertexes of the extended action (1) along with projectors. This is 
justified if one considers the large (= exceeding scale $\mu$) components of 
momenta as very large and the small (= less than $\mu$) components as very 
small. Obviously, the projectors are equal to unit and some vertices are 
forbidden by momentum conservation for such momenta.
The resulting action, $S_{eff}(A^\Gamma)$, allows one to reproduce 
correlators of the
underlying Yang--Mills theory in the Regge limit:
\begin{equation}
\label{limit}
\lim_{\lambda\rightarrow\infty}
\langle \Pi_i (R_{\Gamma_i}(\lambda)A)\rangle ^{S_{YM}}=
\lim_{\lambda\rightarrow\infty}
\langle \Pi_i (R_{\Gamma_i}(\lambda)A^{\Gamma_i})
\rangle ^{S_{eff}},
\end{equation}
where the Regge limit is realized by the rescaling operators 
$R_\Gamma\,(\Gamma=S,R,L,H)$. 
For example, left rescaling $R_L(\lambda)$ is just a Lorentz boost making 
$(-)$-components of momenta
large. It acts on vector fields as follows:
\begin{eqnarray}
\label{remain}
R_L(\lambda)A_+(x^+,x^-,x^i)=\lambda A_+
(\lambda x^+,\frac{x^-}{\lambda},x^i),\,
R_L(\lambda)A_-(x^+,x^-,x^i)=\frac{1}{\lambda}A_-
(\lambda x^+,\frac{x^-}{\lambda},x^i),
\nonumber\\
R_L(\lambda)A_i(x^+,x^-,x^i)=A_i
(\lambda x^+,\frac{x^-}{\lambda},x^i).
\end{eqnarray}
Here and below the Lorentz indices denoted by small Latin characters label 
the transverse components of Lorentz tensors: $i=1,2$. 
$\lambda$ scales as invariant energy of the relative motion of right-
and left-movers; $x^\pm=\frac{x^0\pm x^3}{2}$ are longitudinal components 
of the space-time coordinates (the normalization implies that $kx=k_+x^+
+ k_-x^- + k_ix^i$). The right rescaling, $R_R(\lambda)$, acts as a boost
in the opposite direction: $R_R(\lambda)=R_L(\frac{1}{\lambda})$. The remaining
rescalings, $H$ and $S$, should make both longitudinal momentum components
large and small respectively. The scales for the field magnitudes are adjusted
to ensure the finiteness  of the classical 
equations for the rescaled fields of the effective theory in the limit 
$\lambda\rightarrow\infty$. They are as follows:
\begin{eqnarray}
\label{resh}
R_H(\lambda)A_\pm(x^\pm,x^i)=\frac{1}{\lambda} A_\pm
(\frac{ x^\pm}{\lambda},x^i),\,
R_H(\lambda)A_i(x^\pm,x^i)=A_i
(\frac{ x^\pm}{\lambda}, x^i),
\nonumber\\
R_S(\lambda)A_\pm(x^\pm,x^i)=\lambda A_\pm
(\lambda x^\pm,x^i),\,
R_S(\lambda)A_i(x^\pm,x^i)=\lambda^2 A_i
(\lambda x^\pm,x^i).
\end{eqnarray}
To conclude the description of the relation between the effective theory
and the underlying Yang--Mills theory, we note that one should not forget
altogether the cutoffs introduced by projectors in (2) and use them to define
functional integral implicit in the right-hand-side of equation (3). 
We note here that it is the usual practice to ignore cutoffs and to 
consider various effective theories as the local ones until some 
divergence does not enforce the opposite.

Now we wright down $S_{eff}$ explicitly. Recall that it is obtained by 
substituting 
$A=A^S+A^R+A^L+A^H$ (projectors ignored) into $S_{YM}(A)$ and by 
dropping out vertexes forbidden by momentum conservation. 
To make this dropping out, mark each monomial
of fields entering $S_{YM}(\sum_\Gamma A^\Gamma)$ by two indices, 
$(\omega_L, \omega_R)$.
The indices are taking the following values: $\omega_{L,R}=0,1,2$ depending on
the number of legs injecting large left or, correspondingly, right 
(=$(-)$ or $(+)$) 
momentum component in the vertex. For example, vertex $A^HA^S$ mark by 
$(\omega_L=1,\omega_R=1)$. If the number of legs injecting large 
momentum exceeds two, put the value of corresponding index to 
$2$ ($2$ will stand for ``more than one leg'').
Dropping out vertexes forbidden by momentum conservation amounts to
dropping out the vertexes marked by unit value of any index (large momentum is
injected along a single leg and has no way out). Thus, $S_{eff}$ is a sum of 
four components each of which is a collection of vertexes marked by the 
indices of the same value:
\begin{equation}
\label{decomp}
S_{eff}(\{A^\Gamma\})=\sum_{\omega_L=0,2}\,\sum_{\omega_R=0,2}
S_{(\omega_R,\omega_L)}(\{A^\Gamma\}).
\end{equation}
In what follows we denote $S_{(0,0)}$ as $S_{S(oft)}$. It has no large 
longitudinal
momentum components entering its vertexes and depends on $A^S$ only. 
Analogously,
$S_{(2,0)}=S_{L(eft)}$ (depends on $A^L$ and  $A^S$ only), 
$S_{(0,2)}=S_{R(ight)}$
(depends on $A^R$ and $A^S$ only), and $S_{(2,2)}=S_{H(ard)}$ since it involves
large values of both longitudinal momentum components and depends on all 
the fields.
The explicit expressions for the components of the action are as follows:
\begin{eqnarray}
\label{components}
S_H(\{A^\Gamma\})=S_{YM}(\sum_\Gamma A^\Gamma)+S_{YM}(A^S)-S_{YM}(A^R+A^S)-
S_{YM}(A^L+A^S)-\nonumber\\
-(A^H+A^L)^\mu D^\nu(A^R+A^S)F_{\nu\mu}(A^R+A^S)-\nonumber\\
-(A^H+A^R)^\mu D^\nu(A^L+A^S)F_{\nu\mu}(A^L+A^S)+\nonumber\\
+A^{L\mu}D^\nu(A^S)F_{\nu\mu}(A^S)
+A^{R\mu}D^\nu(A^S)F_{\nu\mu}(A^S)+\nonumber\\
+A^{H\mu}D^\nu(A^S)F_{\nu\mu}(A^S)-
[A^{L\nu},A^{R\mu}]F_{\nu\mu}(A^S)-\nonumber\\
-\frac{1}{2}(D^\mu(A^S)A^{L\nu}-D^\nu(A^S)A^{L\mu})
(D_\mu(A^S)A^R_\nu-D_\nu(A^S)A^R_\mu),\nonumber\\
S_L(A^L,A^S)= S_{YM}(A^L+A^S)-S_{YM}(A^S)-
A^{L\mu}D^\nu(A^S)F_{\nu\mu}(A^S),\nonumber\\
S_R(A^R,A^S)= S_{YM}(A^R+A^S)-S_{YM}(A^S)-
A^{R\mu}D^\nu(A^S)F_{\nu\mu}(A^S),\nonumber\\
S_S(A^S)=S_{YM}(A^S).
\end{eqnarray}
Here $S_{YM}(A)=\frac{1}{2}F_{\mu\nu}(A)F^{\mu\nu}(A)$ 
(with the trace over color indices and integration
over space-time implicit), $\nu$ and $\mu$ are Lorentz indices,
$F_{\mu\nu}(A)=\partial_\mu A_\nu-\partial_\nu A_\mu
-[A_\mu,A_\nu]$, and $D^\nu(A)\Phi=\partial^\nu\Phi-[A^\nu,\Phi]$. 

Now we turn to derivation of local symmetries of effective action 
$S_{eff}=S_H+S_L+S_R+S_S$. To this end note that gauge symmetry of 
$S_{YM}$ states that some local functional vanishes.
It is obtained by an infinitesimal gauge transformation of the action 
and depends on gauge field $A$ and on field $\alpha$ parameterizing 
the transformation. In searching the form this
symmetry takes after replacing the Yang--Mills action by the effective 
one, it is natural to substitute all the fields entering the functional
which vanishes due to the gauge invariance by the sum of four independent 
fields corresponding to disparate values of longitudinal momenta and 
drop out the vertexes forbidden by momentum conservation. 
In particular, the single field $\alpha$ parameterizing
the gauge transformation of the Yang--Mills action is to be substituted 
by the sum:
$\alpha=\alpha^H+\alpha^L+\alpha^R+\alpha^S$. In this way one obtains that
some functional of fields $A^\Gamma,\, \alpha^\Gamma$ vanishes: 
$V(\{A^\Gamma,\alpha^\Gamma\})=0$. 
Consequently, it looks plausible that there are at least four independent 
local symmetries of the effective action corresponding to independent 
fields $\alpha^\Gamma$. The only problem in proving this
is to understand how the vanishing functional is connected to the 
effective action.

It can be noticed that the obtained equality can be written as follows:
\begin{equation}
\label{breaking}
V=\sum_{\Gamma} V_{\Gamma}=0,
\end{equation}
where each $V_\Gamma$ is a field variation of $S_\Gamma$:
\begin{equation}
\label{variation}
V_\Gamma=D_\Gamma S_\Gamma
\end{equation}
It is crucial that variations $D_\Gamma$ are {\it different} for different
$\Gamma$'s. Thus, to describe the symmetry as a symmetry with respect to 
variations of fields we need to consider action $S_{eff}$ as a 
multicomponent (vector) quantity:
$S_{eff}=\{S_\Gamma\}$. As the procedure of breaking 
$S_{eff}=\sum_\Gamma S_\Gamma$
into peaces $S_\Gamma$ is uniquely determined above for any 
polynomial of fields respecting momentum conservation,
there is no ambiguity in this representation of the action as a 
multicomponent quantity. To have a possibility of applying the 
transformation several times consecutively, we need to
break each of the variations $V_\Gamma$ into a sum of functionals 
by the rule defined above:
$V_\Gamma=\sum_{\Gamma'}V_{\Gamma'\Gamma}$. It corresponds 
to breaking each variation $D_\Gamma$ into a sum: 
$D_\Gamma=\sum_{\Gamma'}D_{\Gamma'\Gamma}$. Thus, we arrive at the 
following representation of the gauge transformation of any 
multicomponent functional
$\{S_\Gamma\}$:
\begin{equation}
\label{trans}
(DS)_\Gamma=D_{\Gamma\Gamma'}S_{\Gamma'},
\end{equation}
where summation over $\Gamma'$ is implied. Each component of matrix $D$ is
a field variation. It is  homogeneously and linearly dependent of four gauge 
parameters $\{\alpha^\Gamma\}$: $D=\sum_\Gamma D^\Gamma(\alpha^\Gamma)$,
where each $D^\Gamma$ depends only on $\alpha^\Gamma$. Each $D^\Gamma$ is 
a symmetry of the effective action:
\begin{equation}
\label{sym}
D^\Gamma(\alpha)S_{eff}=
\sum_{\Gamma_1}D^\Gamma_{\Gamma_1\Gamma_2}(\alpha)S_{\Gamma_2}=0,
\end{equation}
where summation over $\Gamma_2$ is implied and $\alpha$ is a field taking 
values in the algebra of gauge symmetries of the underlying Yang--Mills theory.

Now we present an explicit form of matrixes $D^\Gamma$ as a decomposition over
the following basis of matrixes four by four:
\begin{equation}
\label{basis}
(e^{\Gamma_1\Gamma_2})_{\Gamma_1'\Gamma_2'}=\delta^{\Gamma_1}_{\Gamma_1'}
\delta^{\Gamma_2}_{\Gamma_2'}.
\end{equation}
In terms of $e^{\Gamma_1\Gamma_2}$ variations $D^\Gamma$ look as follows:
\begin{eqnarray}
\label{variations}
D^S(\alpha)=(-[A^H\delta_H,\alpha]-[A^L\delta_L,\alpha]-[A^R\delta_R,\alpha]+
D(A^S)\alpha\delta_S)e^{HH}+\nonumber\\
+(-[A^L\delta_L,\alpha]+D(A^S)\alpha\delta_S)e^{LL}+\nonumber\\
+(-[A^R\delta_R,\alpha]+D(A^S)\alpha\delta_S)e^{RR}+\nonumber\\
+(D(A^S)\alpha\delta_S)e^{SS},\nonumber\\
D^R(\alpha)=(-[(A^H+A^L)\delta_H,\alpha]+
D(A^R)\alpha\delta_R)e^{HH}
+(D(A^R)\alpha\delta_R)e^{RR}+\nonumber\\
+(-[A^H\delta_L,\alpha]-[A^R\delta_S,\alpha])e^{HL}+
(-[A^R\delta_S,\alpha])e^{RS},\nonumber\\
D^H(\alpha)=(D(A^H)\alpha\delta_H)e^{HH}+(-[A^H\delta_S,\alpha])e^{HS}
\nonumber\\
+(-[(A^H+A^L)\delta_R,\alpha])e^{HR}+(-[(A^H+A^R)\delta_L,\alpha])e^{HL}.
\end{eqnarray}
Here condensed notations for variations of fields are used: For example,
$(D(A)\alpha\delta_S)A^S_\mu=D_\mu(A)\alpha=\partial_\mu\alpha-[A_\mu,\alpha]$
and $[A\delta_L,\alpha]A^L_\mu=[A_\mu,\alpha]$. The missing expression for 
$D^L$ is obtained from the above expression for $D^R$ by the following 
substitutions of indices:
$R\rightarrow L\,,L\rightarrow R$. We checked by explicit calculation that
$S_{eff}$ defined in (6),(7) is invariant under action of the above $D^\Gamma$,
i.e., equation (11) does hold.

The next step is to calculate commutators of $D^\Gamma$'s. It turns out 
that some of the commutators produce new local symmetries of the 
effective action. These symmetries can be alternatively obtained in the 
way similar to the one used above in the derivation
of $D^\Gamma$. To this end one should start with the symmetry 
transformation of the underlying Yang--Mills theory generated by a 
commutator $[\alpha,\beta]$. Because of the presence of two fields, 
$\alpha$ and $\beta$, the procedure of multiplication of fields and
subsequent dropping out of forbidden vertexes differs from the one 
we considered above with the single gauge parameter $\alpha$. 
It is evident from this derivation that further multiplication 
of fields parameterizing the transformation in the underlying
Yang--Mills theory does not produce new local symmetries of the 
effective theory. This conjecture can be proved by direct calculation 
of double commutators of $D^\Gamma$'s: we checked that double comutators 
can be linearly expressed through $D^\Gamma$ and 
$[D^{\Gamma_1},D^{\Gamma_2}]$. This left us with a closed algebra 
consisting of nine generators. They are four $D^\Gamma$'s defined in 
(13) and five additional generators defined as follows:
\begin{eqnarray}
\label{additional}
D^{HH}([\beta,\alpha])=[D^H(\alpha),D^H(\beta)],\nonumber\\
D^{RR}([\beta,\alpha])=[D^R(\alpha),D^R(\beta)],\,
D^{LL}([\beta,\alpha])=[D^L(\alpha),D^L(\beta)],\nonumber\\
D^{HR}([\beta,\alpha])=[D^H(\alpha),D^R(\beta)], 
D^{HL}([\beta,\alpha])=[D^H(\alpha),D^L(\beta)].
\end{eqnarray}

To present explicitly the total algebra of commutators, we need a new 
notation for the superscripts on the above nine generators: the new 
labeling includes two indexes, $\omega_L$ and $\omega_R$, each index 
runs from $0$ to $2$. The values 
of the new indexes are calculated by the values of the old ones as follows:
Each letter $H$ of the old superscripts gives unit contribution to both 
$\omega_L$ and $\omega_R$, each $L$ gives unit contribution to $\omega_L$, 
each $R$ gives unit contribution to $\omega_R$, and any $S$ give no 
contribution to any $\omega$. For example, $D^{HH}=D^{(2,2)}$ 
($\omega_L$ is first from the left), $D^{HR}=D^{(1,2)}$, 
while $D^S=D^{(0,0)}$. In this notations the algebra of local symmetries
of the effective Regge theory
looks as follows:
\begin{equation}
\label{algebra}
[D^{(\omega_L,\omega_R)}(\alpha),D^{(\omega_L',\omega_R')}(\beta)]=
D^{(\omega_L+\omega_L',\omega_R+\omega_R')}([\beta,\alpha]),
\end{equation}
where if $\omega_\Gamma+\omega'_{\Gamma'}>2$, the sum is replaced by $2$,
and $\alpha, \beta$ belong to the algebra of the underlying Yang--Mills theory.

Now we decompose algebra (15) into a semi-direct sum of $su(N)$ (which is the
gauge algebra of the underlying Yang--Mills theory) and commutative subalgeras.
To this end, note that algebra (15) coincides with the gauge algebra
of Yang--Mills theory which gauge group has the following algebra:
\begin{equation}
\label{algebra}
{\cal A}=su(N)\times{\cal A}_L\times{\cal A}_R,
\end{equation}
where each ${\cal A}_\Gamma$ is a commutative algebra generated by three
generating elements $e^\omega_\Gamma\,(\omega=0,1,2)$ with the 
following table of products:
\begin{equation}
\label{table}
e^{\omega_1}_\Gamma e^{\omega_2}_\Gamma=e^{\omega_1+\omega_2}_\Gamma,
\end{equation}
where the sum of $\omega$'s is understood in the same way as in (15).
This table of products is simplified by going over to a new set of generating
elements: $e_\Gamma=e_\Gamma^0$ (the unit element of ${\cal A}_\Gamma$), 
$\xi_\Gamma=e_\Gamma^1-e_\Gamma^2$, and $P_\Gamma=e_\Gamma^2$. Now (17)
takes the following form:
\begin{eqnarray}
\label{newtable}
e_\Gamma e_\Gamma=e_\Gamma,\,e_\Gamma \xi_\Gamma=\xi_\Gamma,\,
e_\Gamma P_\Gamma=P_\Gamma,\nonumber\\
\xi_\Gamma\xi_\Gamma=0,\,\xi_\Gamma P_\Gamma=0,\nonumber\\
P_\Gamma P_\Gamma=P_\Gamma.
\end{eqnarray}

From table (18) it is evident that  sets $su(N)\times P_L\times P_R$,
$su(N)\times \xi_L\times \xi_R$, $su(N)\times P_L\times \xi_R$, and
$su(N)\times \xi_L\times P_R$ generate ideal ${\cal I}$ of algebra ${\cal A}$.
This ideal can be decomposed into a direct sum of subalgebras corresponding to
the above sets. Factoralgebra ${\cal A/I}$ has in turn an ideal
generated by $su(N)\times P_L\times e_R$, $su(N)\times e_L\times P_R$,
$su(N)\times \xi_L\times e_R$, and $su(N)\times e_L\times \xi_R$. Thus, we
have the following decomposition of ${\cal A}$:
\begin{equation}
\label{decomposition}
{\cal A}=[su_S(N)\uplus(su_L(N)\oplus su_R(N)\oplus A_L\oplus A_R)]\uplus
(su_H(N)\oplus A_H\oplus A_{HL}\oplus A_{HR}),
\end{equation}
where $\uplus$ denotes semi-direct sum of algebras, $\oplus$ 
denotes direct sum of algebras, and
\begin{eqnarray}
\label{pieces}
su_S(N)=su(N)\times e_L\times e_R,\, 
su_L(N)=su(N)\times P_L\times e_R,\nonumber\\
su_R(N)=su(N)\times e_L\times P_R,\,su_H(N)=su(N)\times P_L\times P_R,
\end{eqnarray}
while the Abelian subalgebras $A_L$, $A_R$,  $A_H$, $A_{HL}$ and 
$A_{HR}$ are as follows:
\begin{eqnarray}
\label{abelian}
A_L=su(N)\times \xi_L\times e_R, \,A_R=su(N)\times e_L\times \xi_R,\nonumber\\ 
A_H=su(N)\times \xi_L\times \xi_R,\nonumber\\
 A_{HL}=su(N)\times \xi_L\times P_R,\, A_{HR}=su(N)\times P_L\times \xi_R.
\end{eqnarray}
Dimension of each Abelian algebra
entering the decomposition (\ref{decomposition}) coincides with the one of
the gauge algebra of the underlying Yang--Mills theory ($N^2-1$ for $su(N)$).

The fact that there are Abelian algebras in the decomposition 
(\ref{decomposition}) makes it plausible that corresponding group 
is not a simply connected space. This in turn makes plausible the 
existence of solitons in the effective Regge theory.
This guess finds some grounds in a direct analysis of the 
classical field equations
of the effective Regge theory which will be published elsewhere.

In conclusion, effective action to describe Regge limit of Yang--Mills theories
is introduced, its local symmeties are determined, and a decomposition of the
 algebra of symmetries into a semi-direct sum of Abelian algebras and
four copies of the gauge algebra of the underlying Yang--Mills theory 
is found. The local symmetries are defined as transformations of any 
translationally invariant field functional but cannot be defined
as field transformations. However, the latter is possible by the expense of
introducing a notion of multicomponent action. A conjecture is risen
on the presence of solitons in the effective Regge Yang--Mills theory.

We thank V.T. Kim, V.A. Kuzmin and A.M. Boyarsky, and the staff of the 
Theory Division of INR (Moscow) for stimulating discussions.
This work was supported in part by the Russian Foundation
for Basic Research, grants No. 96-02-16717 and 96-02-18897.



\begin{thebibliography}{10}

\bibitem{Ver}
H. Verlinde and E. Verlinde, 
QCD at High-Enrgies and Two-Dimensional Field Theory,
hep-th/9302104 
\bibitem{Ar}
I.Ya. Aref'eva, Phys.Lett.B325 (1994) 171

\bibitem{Lip}

L.N. Lipatov, Nucl.Phys.B452 (1995) 369 

\bibitem{KP}
V.T. Kim and G.B. Pivovarov,
Phys.Rev.Lett.79 (1997) 809 

\end{thebibliography}
\end{document}